\documentclass[aps,prl,reprint,twocolumn,floatfix,superscriptaddress,longbibliography]{revtex4-1}
\usepackage{amsfonts, amsmath, amssymb}
\usepackage{graphicx}
\usepackage{dcolumn}
\usepackage{bm}
\usepackage[normalem]{ulem}
\usepackage{color}
\usepackage[utf8]{inputenc}
\usepackage[colorlinks,citecolor=blue,linkcolor=blue,urlcolor=blue]{hyperref}
\usepackage{braket}
\usepackage[caption=false]{subfig}
\usepackage{color}
\usepackage{soul}
\usepackage{mathtools}
\usepackage{float}
\newcommand{\be}{\begin{equation}}
\newcommand{\ee}{\end{equation}}
\newcommand{\ba}{\begin{eqnarray}}
\newcommand{\ea}{\end{eqnarray}}

\newcommand{\bdag}{b^{\dagger}}

\include{physics}

  {\left\lbrace\begin{array}{@{}l@{}}}%
  {\end{array}\right.}

\begin{document}

\title{Entanglement dynamics in dissipative photonic Mott insulators}

\author{Kaelan Donatella}
\affiliation{Universit\'{e} de Paris, Laboratoire Mat\'{e}riaux et Ph\'{e}nom\`{e}nes Quantiques, CNRS-UMR7162, 75013 Paris, France}

\author{Alberto Biella}
\affiliation{JEIP, USR 3573 CNRS, Collège de France, PSL Research University, 11 Place Marcelin Berthelot, 75321 Paris Cedex 05, France}
\affiliation{Universit\'{e} de Paris, Laboratoire Mat\'{e}riaux et Ph\'{e}nom\`{e}nes Quantiques, CNRS-UMR7162, 75013 Paris, France}

\author{Alexandre Le Boit\'e}
\affiliation{Universit\'{e} de Paris, Laboratoire Mat\'{e}riaux et Ph\'{e}nom\`{e}nes Quantiques, CNRS-UMR7162, 75013 Paris, France}

\author{Cristiano Ciuti}
\affiliation{Universit\'{e} de Paris, Laboratoire Mat\'{e}riaux et Ph\'{e}nom\`{e}nes Quantiques, CNRS-UMR7162, 75013 Paris, France}

\begin{abstract}
We theoretically investigate the  entanglement dynamics in photonic Mott insulators in the presence of particle losses and dephasing. We explore two configurations where entanglement is generated following the injection or extraction of a photon in the central site of a chain of cavity resonators. We study the  entanglement negativity of two-site reduced density matrices as a function of time and inter-site distance. Our findings show that in spite of particle losses the quantum entanglement propagation exhibits a ballistic character with propagation speeds related to the differerent quasiparticles that are involved in the dynamics, namely photonic doublons and holons respectively. Our analysis reveals that photon dissipation has a strikingly asymmetric behavior in the two configurations with a much more dramatic role on the holon entanglement propagation than for the doublon case. 
\end{abstract}

\date{\today}
\maketitle

After being a subject of early intense debate at the dawn of quantum mechanics \cite{Schrodinger:1935, Einstein:1935}, entanglement is now recognized as a key feature of quantum physics \cite{Horodecki:2009}. The efforts towards building a complete mathematical description of this notion were instrumental in the development of quantum information.  In this context, the core of the theory is centered around three main tasks: detecting~\cite{Guhne:2009}, quantifying~\cite{Plenio:2007} and manipulating entanglement~\cite{Bennett:1996}. The progress made on these three fronts would allow to outperform  classical methods in the fields of metrology~\cite{Giovannetti:2011}, cryptography~\cite{Gisin:2002} and  computation~\cite{Nielsen:2000}.

In addition to providing sound foundations to the field of quantum information, entanglement theory has also paved the way to new discoveries in other areas of physics. As anticipated at the beginning of the millenium~\cite{Preskill:2000}, quantities such as the entanglement entropy have proved to be very valuable tools for characterizing the ground-state wave function of many-body quantum systems~\cite{Amico:2008, DeChiara:2018, Islam:2015}. The study of entanglement in many-body systems has not been restricted to their ground state properties: entanglement dynamics and its propagation in space in quantum systems has also been the subject of intense research activities for spin chains \cite{amicospins, Jurcevic:2014, Kastner:2015}, fermionic \cite{alba2020spreading} and bosonic systems \cite{Lauchli:2008, cheneau_light-cone-like_2012, Daley:2012, bernier_light-cone_2018}.

The understanding of entanglement in open quantum many-body systems represents a timely frontier of research \cite{Aolita_2015} that is of fundamental importance because much less is known with respect to the state-of-the-art in isolated quantum many-body systems at thermal equilibrium or exhibiting unitary Hamiltonian dynamics. Whereas in general experimentalists try to protect their system from interacting with its environment, other approaches based on the general concept of "reservoir engineering" try to exploit the openness of a system and take advantage of judiciously designed dissipation to reach non-trivial quantum states in the transient regime \cite{Leghtas853} or in the steady state \cite{Mott}. In recent years, experimental progress in tailoring effective photon-photon interactions in cavity and circuit quantum electrodynamics (QED) devices has lead to the emergence of controllable quantum optical many-body systems \cite{rmpciuti,Schmidt:2013,HartmannRev2016, AngelakisRev2016,Roy:2017}. Unlike most condensed matter setups where the system is close to thermal equilibrium, this new class of systems are open quantum platforms in which intrinsic losses, due to the photon finite lifetime, have to be compensated by an external coherent or incoherent driving.  

Although several works have been devoted to transport properties of strongly-correlated photonic platforms~\cite{Biella:2015, Lee:2015, Mertz:2016, Debnath:2017}, entanglement and correlation propagation in driven-dissipative systems have  been studied for dissipative free fermion systems  \cite{alba2020spreading} and remain largely unexplored.
The recent experimental demonstration of dissipatively stabilized photonic Mott insulators \cite{Mott} in chains of superconducting microwave resonators paves the way to the exploration of such an exciting frontier.

        In this Letter, we theoretically explore the physics of entanglement propagation in photonic Mott insulators, showing genuine physical effects associated to the openness of such systems. In contrast to most works about correlation propagation in interacting bosonic systems, here we do not consider global quenches of the system that typically consist in abruptly changing the value of the interaction strength in all the lattice \cite{cheneau_light-cone-like_2012,  bernier_light-cone_2018}. Instead, we consider two configurations where one photon is injected or removed from one cavity in the middle of a chain and investigate the propagation of entanglement that is produced between distant sites as a function of time and of their spatial separation. Such a study is achieved by monitoring the negativity of two-site reduced density matrices, that witnesses entanglement. We show a strinking different role of photon dissipation in the two configurations. 

\begin{figure}[t!]
    \centering
    \includegraphics[scale=0.4]{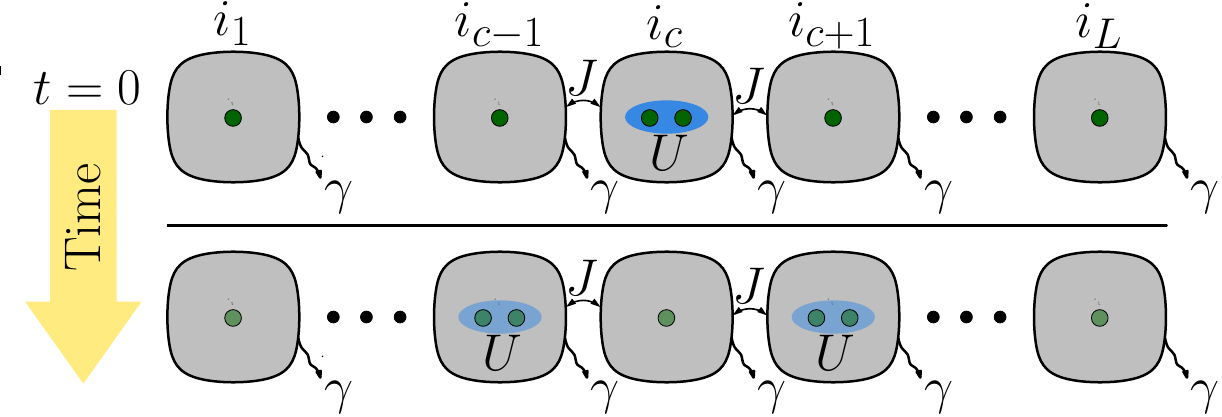} \\
    \includegraphics[scale=0.55]{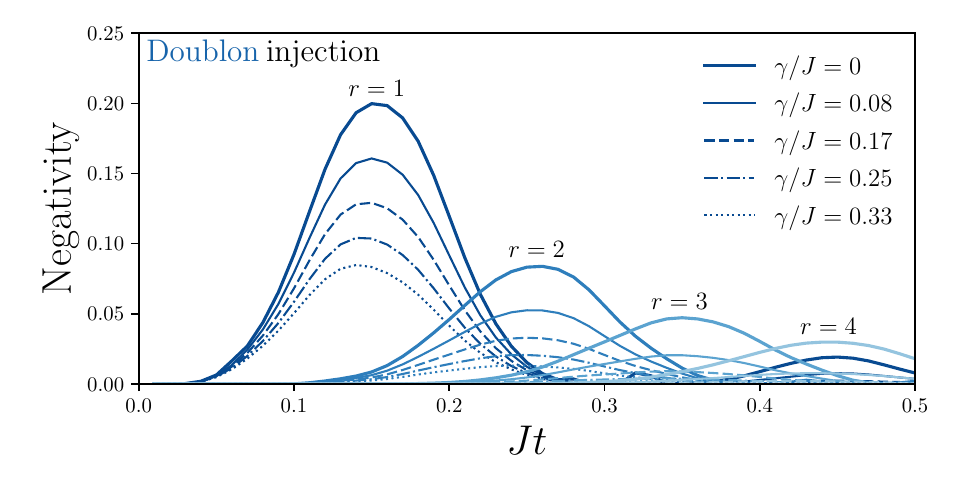}
    \caption{Upper panel: sketch of the considered system, a chain of coupled photonic resonators with on-site photon-photon interaction $U $ and nearest-neighbor hopping coupling $J$. The top chain depicts the initial time configuration with a Mott insulator of photons (one photon per cavity) where a double occupation (doublon) has been created in the central site $i=i_c$. The bottom chain depicts the configuration at a later observation time $t$, with entanglement existing between distant sites. The photonic modes are subject to losses and dephasing. Lower panel: entanglement negativity $\mathcal {N}_r(t)$ between sites $i_{c-r}$ and $i_{c+r}$ as a function of time $t$ for different values of the spatial separation $r=\{1,2,3,4\}$ from the central site $i_c$. The shade of the lines gradually decreases going from $r=1$ to $r=4$. Calculations were performed via MPO simulations  (bond link dimension $\chi=200$) on a chain of $L = 20$ cavity sites. For each value of $r$, results for different values of the photon loss rate $\gamma$ in units of the hopping $J$ are shown. The initial state at $t=0$ is $\ket{\Psi_D}$ (see the text) corresponding to a doublon excitation localized in the central site. In this figure, the pure dephasing rate $\Gamma_d$ is 0. The on-site interaction for all the cavities is $U/J=33.3$.}
    \label{DoublonPropagation}
\end{figure}
\textit{Model - }Let us consider a chain of $L$  coupled nonlinear electromagnetic resonators described by a Bose-Hubbard Hamiltonian:
\begin{gather}
     \mathcal{H} =\sum_{i=1}^L\left(\omega_cb^{\dagger}_{i}b_i+\frac{U}{2}b^{\dagger}_{i}b^{\dagger}_{i}b_{i}b_{i}\right)-J \sum_{i=1}^{L-1} (b^{\dagger}_{i} b_{i+1} + h.c.),
\end{gather}with $\omega_c$ the cavity mode frequency, $U$ the photon-photon (Kerr) on-site interaction, $J$ the nearest-neighbor photon hopping coupling, and $b_i$ ($b_i^{\dagger}$) the annihilation (creation) photon operators for each site. The physical systems described by the Bose-Hubbard Hamiltonian include, but are not limited to, lattices of microwave resonators in circuit QED platforms \cite{Houck2012, rmpciuti, cQEDTransition, HartmannRev2016, AngelakisRev2016, Carusotto2020},  semiconductor microcavities \cite{rmpciuti, bloch} and ultracold gases in optical lattices \cite{Greiner2002,PhysRevLett.81.3108}. These systems exhibit dissipation and dephasing due to the coupling to the environment. In cold atom systems dephasing is dominant \cite{bernier_light-cone_2018} while for microwave photons in circuit QED platforms particle loss is typically the most important channel \cite{cold_dephasing, Hartmann_2016}. Within an open quantum systems approach, the time evolution of the system density matrix $\rho$ can be described by the following Lindblad master equation \cite{BRE02}:
\begin{equation}
\frac{d\rho}{dt}=-{\rm i} [\mathcal{H},\rho]+\frac{1}{2}\sum_{i=1}^L \sum_{{\mathcal C}} 2J^{({\mathcal C}) \dagger}_i\rho J^{({\mathcal C})}_i-\{J^{({\mathcal C})\dagger}_i J^{({\mathcal C})}_i,\rho \},
\end{equation} with $J^{({\mathcal C})}_i$ the jump operator for the $i$-th site and the dissipation channel ${\mathcal C}$. When the temperature is low enough and the thermal photon occupancy is negligible,  the jump operator for the particle loss channel (${\mathcal C} = l$) due to the finite photon lifetime reads $J^{(l)}_i =\sqrt{\gamma} b_i$, where $\gamma$ is the photon loss rate. The pure dephasing channel (${\mathcal C} = d$) due to fluctuations in the environment is described by the jump operator $J^{(d)}_i = \sqrt{2\Gamma_d}\bdag_ib_i$, with $\Gamma_d$ the pure dephasing rate.

In this work we will focus on the strongly correlated limit $U\gg J$.  In such a regime, in order to describe the physics of a photonic Mott insulator with one photon per site, we can safely truncate the local Hilbert space to $2$ photons per site by retaining only the $\vert 0 \rangle$, $\vert 1 \rangle$, and $\vert 2 \rangle$ Fock number states. The validity of this assumption was carefully tested numerically by increasing the local Hilbert space cutoff. A Mott insulator phase corresponding to one photon per site for $U \gg J$ is approximately described by the factorized state \begin{equation}
\label{Mott_eq}
    \ket{\Psi_{\text{Mott}}}=\ket{1}_1\otimes\ket{1}_2\otimes...\otimes\ket{1}_L=\ket{11...1}.
\end{equation}In the regime of strong interactions, the Hamiltonian can be diagonalized by using generalized Jordan-Wigner and Bogoliubov transformations \cite{cheneau_light-cone-like_2012}, via a mapping  to a spin-1 model. This leads to a quasiparticle picture containing two types of fermionic-like excitations: doublons and holons. The ground state (quasiparticle vacuum) corresponds to the Mott insulator $\ket{\Psi_{\text{Mott}}}$ on top of which quasiparticles propagate. These fermionic quasiparticles are described by local creation operators $d^{\dagger}_i$ and $h^{\dagger}_i$ for doublons and holons respectively, such that  $d^{\dagger}_i\ket{1}_i=\ket{2}_i, h^{\dagger}_i \ket{1}_i=\ket{0}_i$. 
 
\begin{figure}[t!]
    \centering
    \includegraphics[scale=0.4]{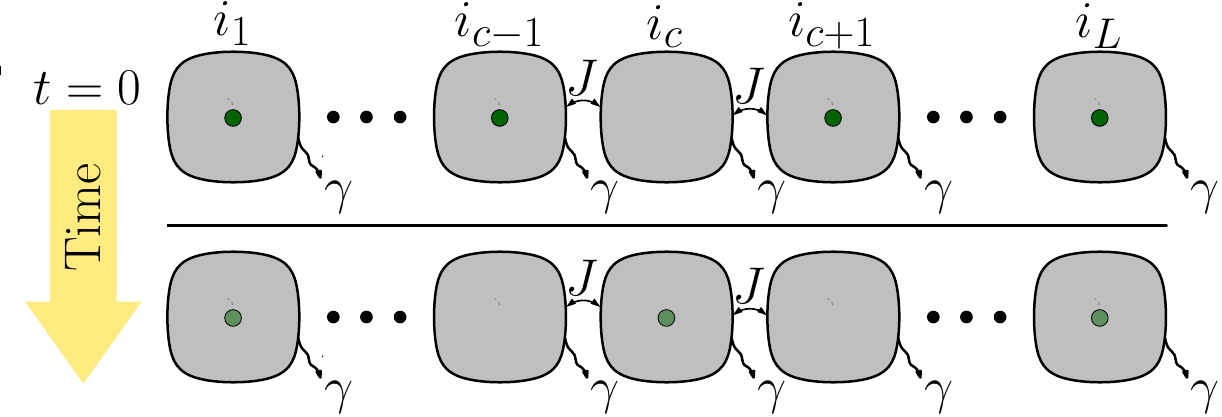} \\
    \includegraphics[scale=0.55]{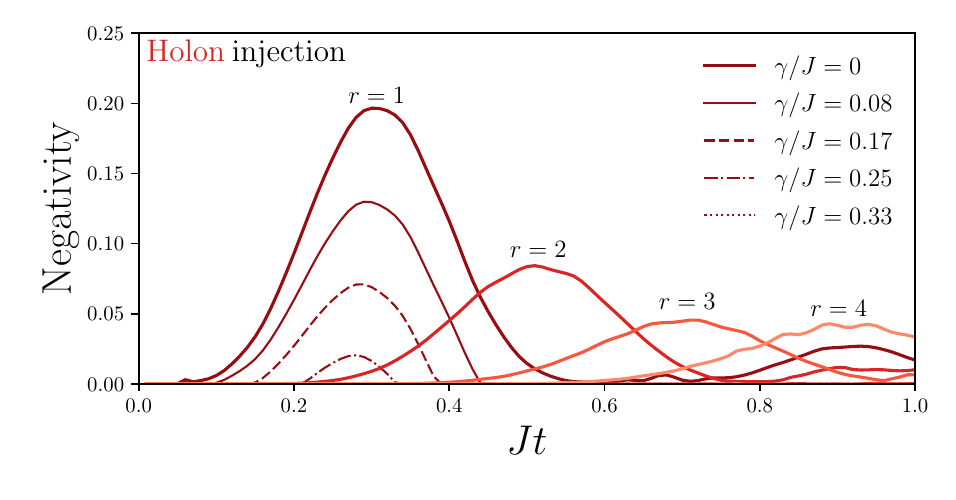}
    \caption{Upper panel: sketch like in Fig. \ref{DoublonPropagation}, but where the initial state has an empty central site (holon). Lower panel:
    temporal dynamics of the  negativity with same parameters as in Fig. \ref{DoublonPropagation}, but  with an initial state $\vert \Psi_H \rangle$ corresponding to a holon localized in the central site.  }
    \label{Hole_Propagation}
\end{figure}

 \textit{Entanglement generation protocol - }
 Since $\vert {\Psi_{\text{Mott}}}\rangle $ is a factorized state, an interesting question is how to perturb such a
 photonic Mott insulator in order to create entanglement in a simple way and study its propagation in a direct fashion. In the following we will show that this is possible by injecting (or removing) one photon from an occupied site. As shown in the upper panels of Figs. \ref{DoublonPropagation} and \ref{Hole_Propagation}, we will consider such manipulation on the central site of a linear chain of resonators. In the case of a photonic insulator with a large $U$, this can be achieved simply by applying a coherent $\pi$-pulse drive on the central site that induces a Rabi rotation from the $\vert 1 \rangle$ to the $\vert 2 \rangle$ (or to the $\vert 0\rangle$) Fock number state in the considered site. We have explicitly verified that such operation can be performed with fidelity close to $1$ thanks to the strong anharmonicity produced by the large on-site interaction $U$.
 This way, it is possible to prepare the state $
    \ket{\Psi_D}=\ket{1...2...1}$ ($
    \ket{\Psi_H}=\ket{1...0...1}$) where $D$ ($H$) stands for doublon (holon), corresponding to the injection of a single localized  excitation on top of the quasiparticle vacuum. In the doublon case, the excess photon in the central site can hop to the right nearest-neighbor site or, with the same probability, to the left site. Due to the symmetry of the chain with respect to the central site and the lack of which-path information, such propagation creates an entangled state that can propagate along the chain. For circuit QED platforms, a Mott insulator can be prepared and maintained through an active stabilization process \cite{Mott}. In the following, we will consider the dissipative dynamics of the system after the creation of the localized doublon (holon) in the absence of stabilization.

\textit{Entanglement detection - }In order to witness bipartite entanglement between two partitions $A$ and $B$ of the system, we have considered the negativity function \cite{negativity}, defined as $ \mathcal{N}(\rho_{AB}) = \sum_{\lambda < 0} \vert \lambda \vert $, where the sum is taken over the negative eigenvalues $\lambda$ of $\rho^{\Gamma_A}_{AB}$, which is the partial transpose with respect to subsystem $A$ of the joint density matrix $\rho_{AB}$. In the following,   $A$ and $B$ will be two resonators at a distance $r$ from the central cavity ($i = i_{c}$). The time-dependent negativity of the reduced density matrix for these two sites at the positions $i_{c-r}$ and $i_{c+r}$ will be denoted by $\mathcal{N}_r(t)$. For systems with a relatively moderate Hilbert space dimension, we can compute the time evolution of the full density matrix of the system via an exact integration of the master equation. Once we get the full density matrix, we can trace out with respect to the degrees of freedom of all the sites except the two sites under study.

In the regime where we can consider only the $\vert 0 \rangle$, $\vert 1 \rangle$, and $\vert 2 \rangle$ states as local basis for a given site, we can reconstruct a two-site reduced density matrix by exploiting the fact that any Hermitian operator can be decomposed over the generators of the group associated to its Hilbert space \cite{tomography}. As we truncated the local Hilbert space to states with up to $2$ photons, the generators of the SU(3)$\otimes$ SU(3) group allows us to reconstruct the reduced density matrix as:
\begin{equation}
\rho_3^{(2)}=\frac{1}{9}\sum_{i_1,i_2=0}^{8}r_{i_1i_2} \Lambda^{(i_1)}\otimes \Lambda^{(i_2)},
\end{equation} with $\Lambda^{(i)}$ the generators \footnote{The nonzero matrix elements of the Hermitian $\Lambda^{(i)}$ matrices, such that $\Lambda^{(i)}_{r,s} = (\Lambda^{(i)}_{s,r})^{\star}$, are the following ones:   $\Lambda^{(0)}_{11} =  \Lambda^{(0)}_{22} =\Lambda^{(0)}_{33} = 1$, $ \Lambda^{(1)}_{12} = 1$, 
$\Lambda^{(2)}_{12} = - i$, $\Lambda^{(3)}_{11} = - \Lambda^{(3)}_{22} = 1$, $\Lambda^{(4)}_{13} = 1$,$\Lambda^{(5)}_{13} = -i$, $\Lambda^{(6)}_{23} =1$,
$\Lambda^{(7)}_{23} = -i$, $\Lambda^{(8)}_{22} = 1/\sqrt{3}$ and $\Lambda^{(8)}_{33} = -2/\sqrt{3}$. 
} of the SU(3) group and \begin{gather*}
    r_{i_1i_2}=\frac{9 \langle \Lambda^{(i_1)}\otimes\Lambda^{(i_2)} \rangle}{\text{tr}  ( (\Lambda^{(i_1)}\otimes\Lambda^{(i_1)})^2 )}.
\end{gather*} 

Using this method we can reconstruct the two-site density matrix for all times and obtain the entanglement negativity. 
Since the Hermitian operators $\Lambda^{(i)}$ can also be expressed as a function of the bosonic creation and annihilation operators up to their third-order power \footnote{The eight generators $\Lambda^{(i)}$ can be expressed in terms of the bosonic annihilation and creation operators. Namely: $\Lambda^{(1)} =\frac{1}{2}(b b^{\dagger 2}+b^{2} b^{\dagger})$, $\Lambda^{(2)}=\frac{i}{2}(b^2 b^{\dagger}-b b^{\dagger 2})$, $\Lambda^{(3)}=\frac{3}{4}b^{2} b^{\dagger 2}-\frac{1}{2}b b^{\dagger}$, $\Lambda^{(4)}=\frac{1}{\sqrt{2}}(b^{\dagger 2}+b^2)$, $\Lambda^{(5)}=\frac{i}{\sqrt{2}}(b^2 - b^{\dagger 2})$, $\Lambda^{(6)}=\frac{1}{\sqrt{2}}(b^{\dagger 2} b+b^{\dagger} b^2)$,
$\Lambda^{(7)}=\frac{i}{\sqrt{2}}(b^{\dagger} b^{2} -b^{\dagger 2} b)$, and $\Lambda^{(8)} =\frac{1}{\sqrt{3}}(bb^{\dagger}-b^{\dagger} b)$.}, this tomographic method can be used in experiments to measure the two-site entanglement negativity. Indeed, the measurement of expectation values of moments of the photon fields has become a rather standard procedure in circuit QED platforms (see, e.g. \cite{walraff}). Note that we have conveniently developed and used this approach for numerical simulations based on the Matrix Product Operator (MPO) technique  \cite{mpo2, mpo3,Biella:2015}. Indeed, MPO simulations are effective to simulate longer chains of cavities but do not allow for a direct access to the full system density matrix, an issue that we bypassed with the procedure described above.

\textit{Results and discussion - }
In Fig. \ref{DoublonPropagation}, we report results for the negativity ${\mathcal N}_r(t)$  for different values of $r$ and of the photon loss rate $\gamma$ (here no pure dephasing is considered, $\Gamma_d = 0$). The negativity shows a well resolved peak for most values of the spatial separation $r$ and $\gamma/J$: increasing $r$ delays the negativity peak, showing a clear entanglement propagation. A revival peak of entanglement is visible in the $r=1$ curve at longer times \footnote{Previous studies of two-qubit systems with non-Markovian environments have revealed entanglement revival effects \cite{revival, revival2}. In our system, the two-site dynamics is non-unitary and non-Markovian even for $\gamma = 0$ since the other sites of the chain have been traced out for the calculation of the negativity.}. The value of the negativity peaks decreases with increasing dissipation $\gamma$. However, it is remarkable that the entanglement propagation speed is negligibly influenced by dissipation and remains essentially ballistic. In Fig. \ref{Hole_Propagation}, we report the analogous dynamics of negativity for the other configuration where the holon state $\vert \Psi_H \rangle$ is prepared. With the same  parameters as in Fig. \ref{DoublonPropagation},  in the holon case not only the propagation speed is slower, but the role of dissipation is more dramatic, as we do not see any peaks for the chosen values of $\gamma/J$ as soon as $r>1$. 

\begin{figure}[t!]
    \centering
    \includegraphics[scale=0.55]{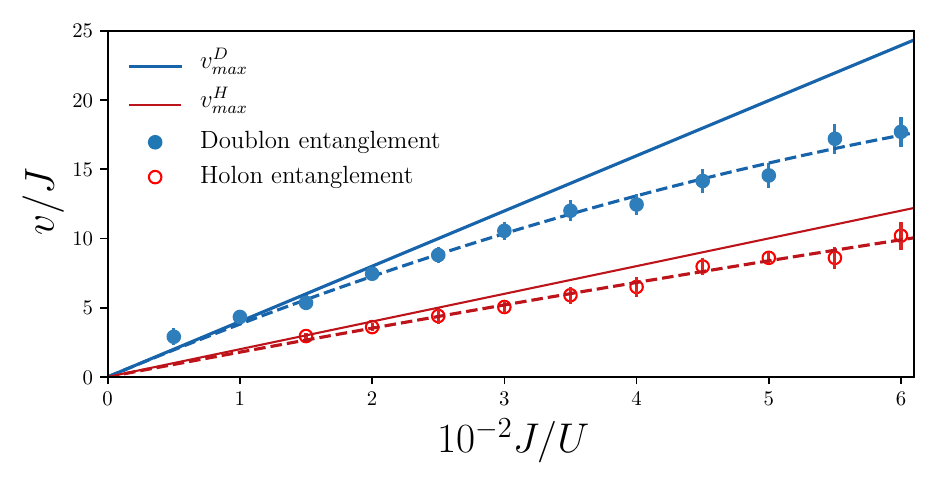}
    \caption{Entanglement propagation speed (units of $J$) versus the normalized hopping $J/U$. Filled (blue) circles: entanglement speed for the configuration corresponding to the injection of one additional photon (doublon injection). Empty (red) circles: entanglement speed for the holon injection. Solid (blue) thick line: maximal  speed for doublon quasiparticles in a closed Hamiltonian system. Thin (red) solid line: maximal holon speed. Dashed lines are polynomial fits with linear and quadratic terms in $J/U$. Error bars were estimated taking into account uncertainty due to time discretization and to the finite bond link dimension in the MPO calculations. Parameters: $\gamma/J=0.1$ and for the doublon (holon) configuration $U/\gamma = 100$ ($1000$).}  
    \label{Speed}
\end{figure}

\begin{figure}[t!]
    \centering
    \includegraphics[scale=0.57]{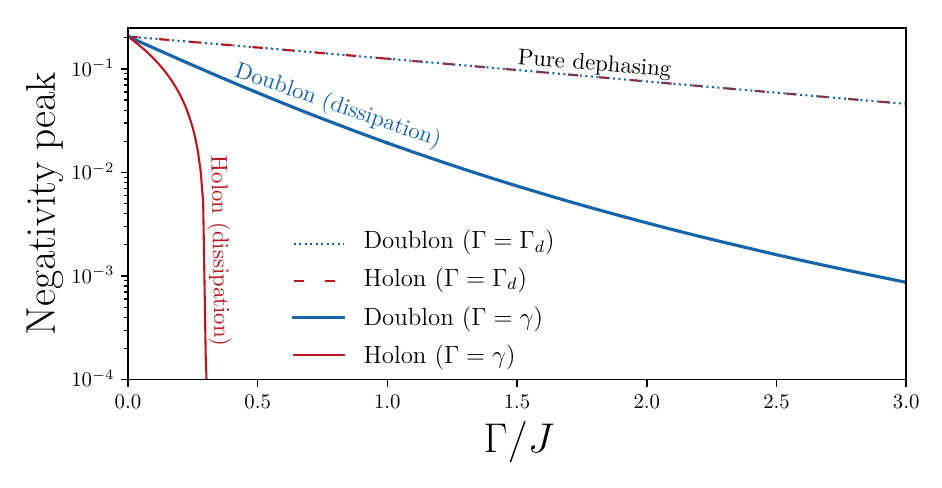}
    \caption{Peak value of ${\mathcal{N}_{r=1}}$ (log scale) as a function of the photon dissipation rate $\gamma$ or dephasing $\Gamma_d$. Thick (thin) solid line: curve for the doublon (holon) configuration versus $\gamma$ when only photon particle losses are present ($\Gamma_d = 0$). The holon injection case has been calculated for a chain with a hopping coupling $2J$  to have the same maximal speed of a doublon in a chain with hopping coupling $J$ (see text). Dotted (doublon) and dashed (holon) lines: negativity peak versus $\Gamma_d$ when there is only the  pure dephasing channel ($\gamma =0$).   Parameters{\tiny }: $L=5, U/J=100.$}
    \label{Asymmetry}
\end{figure}

  In Fig. \ref{Speed}, we report  the calculated entanglement propagation speeds versus $J/U$ both for the case of photon injection  (doublon excitation) and  extraction (holon excitation). In the same plot, we have also reported the maximal propagation speed of doublons and that of holons for a closed system, namely
$  v^D_{max}=4J\left[1-\frac{4J^2}{U^2}\right]+\mathcal{O}\left(\frac{J^3}{U^4}\right),
$ and
$
  v^H_{max}=2J\left[1+  \frac{17J^2}{2U^2}\right]+\mathcal{O}\left(\frac{J^3}{U^4}\right)$.
The fact that the speed is negligibly altered by the dissipation is at first surprising because in the quasiparticle picture the injected doublon (holon) propagates on top of a dissipating background of photons. In the presence of losses, the probability of having such photonic Mott insulator background decreases with time, reducing the entanglement peak, but not the associated doublon (holon) speed. 
 The effect of dissipation for the two considered configurations  is presented in Fig. \ref{Asymmetry} in which the peak value of $\mathcal{N}_{r=1}$ is plotted.  To compare the genuine effect of dissipation and dephasing, we considered a holon propagation in a chain with hopping coupling $2J$ and a doublon propagation in a chain with hopping $J$ in order to have the same speed (for $U/J \gg 1 $ the speeds differ by a factor $2$). From the peak value of the negativity (occurring at the same time),  we see that pure dephasing acts on the two cases in the same identical way (dashed lines), with an exponential decay of the negativity. Indeed, pure dephasing conserves the total number of particles and does not break the particle-hole symmetry. On the other hand, in the presence of photon losses, our investigation reveals a striking asymmetry between the doublon (thick solid line) and holon (thin solid line) cases. Indeed,  the negativity vanishes much faster for the holon configuration even when the speed is the same. The asymmetry can be qualitatively understood if we consider the quantum jump picture associated to the photon loss channel and the two-site reduced density matrix. In the holon case, by diagonalizing such reduced  density matrix, we have found that the entanglement is mostly due to the Bell state $\vert \psi_{H, +}\rangle = \frac{1}{\sqrt{2}} \left (\vert 0\rangle_{i_{c-r}}\vert 1\rangle_{i_{c+r}} + \vert 1\rangle_{i_{c-r}}\vert 0\rangle_{i_{c+r}} \right )$. In a single quantum trajectory picture, a single quantum jump due to photon decay in one of the two sites transforms such state into the factorized state $\vert 0\rangle_{i_{c-r}} \otimes \vert 0\rangle_{i_{c+r}}$. 
By contrast, in the doublon configuration, the entanglement is mostly due to the state
$\vert \psi_{D, +}\rangle = \frac{1}{\sqrt{2}} \left (\vert 2\rangle_{i_{c-r}}\vert 1\rangle_{i_{c+r}} + \vert 1\rangle_{i_{c-r}}\vert 2\rangle_{i_{c+r}} \right )$.
In this case, the quantum jump produced by a photon loss in site $i_{c-r}$ produces the (normalized) state $ \frac{1}{\sqrt{3}}  \left (\sqrt{2} \vert  1\rangle_{i_{c-r}}\vert 1\rangle_{i_{c+r}} + \vert 0\rangle_{i_{c-r}}\vert 2\rangle_{i_{c+r}} \right )$, which is still entangled. A photon loss in site $i_{c+r}$ produces an analogous state. The quantitative results in Fig. \ref{Asymmetry} show a remarkable non-exponential dependance of both doublon and holon negativity as a function of the dissipation rate. The holon negativity becomes exactly zero for a finite value of the dissipation $\gamma$ whereas the decay  in the doublon case slows down for increasing values of $\gamma$. Indeed, the  two-site reduced density matrix satisfies an effective master equation, which is in general non-markovian  as  obtained by tracing out the other degrees of freedom of the chain, whose dynamics is non-trivial.  

\textit{Conclusion - } In this letter, we have theoretically explored the physics of entanglement propagation in photonic Mott insulators in the presence of photon particle losses and dephasing. We have investigated a scheme where the entanglement is generated by injecting (or extracting) a photon from a site of a photonic Mott insulator. We have introduced a relatively simple quantum state tomography protocol, valid in the limit of strong photon-photon interactions, to study the bipartite entanglement properties. Our scheme is particularly suited for circuit QED platforms exhibiting strong photon-photon interactions and allowing the measurement of quantum optical correlation functions between distant sites. In spite of the losses, the propagation of the negativity peak exhibits a speed, which is close to the doublon (holon) quasiparticle propagation speed respectively in the case of the injection (extraction) of a photon. Remarkably, the impact of particle losses is highly asymmetric for these two configurations, while pure dephasing does not break the doublon-holon symmetry. Our work paves the way to new investigations on the entanglement propagation in open quantum systems. A future interesting research direction is the characterization of the entanglement dynamics at long times (diffusive vs. ballistic) and the quest for universal features underlying the dynamics of correlations in this class of systems. Another challenging problem to be investigated in the future, given the recent experimental success in the dissipative stabilization of photonic Mott insulators \cite{stabilizer}, is the search for protocols to stabilize entanglement propagation  in open quantum systems.

\acknowledgements{
We would like to acknowledge stimulating discussions with Davide Rossini. This  work  was supported by the ERC project CORPHO (no. 616233), the H2020-FETFLAG project PhoQus (no.  820392) and by project ANR UNIQ DS078. We would like to sincerely thank all the essential workers for their invaluable role during the recent pandemic crisis.}

\bibliography{bib}

\begin{thebibliography}{53}%
\makeatletter
\providecommand \@ifxundefined [1]{%
 \@ifx{#1\undefined}
}%
\providecommand \@ifnum [1]{%
 \ifnum #1\expandafter \@firstoftwo
 \else \expandafter \@secondoftwo
 \fi
}%
\providecommand \@ifx [1]{%
 \ifx #1\expandafter \@firstoftwo
 \else \expandafter \@secondoftwo
 \fi
}%
\providecommand \natexlab [1]{#1}%
\providecommand \enquote  [1]{``#1''}%
\providecommand \bibnamefont  [1]{#1}%
\providecommand \bibfnamefont [1]{#1}%
\providecommand \citenamefont [1]{#1}%
\providecommand \href@noop [0]{\@secondoftwo}%
\providecommand \href [0]{\begingroup \@sanitize@url \@href}%
\providecommand \@href[1]{\@@startlink{#1}\@@href}%
\providecommand \@@href[1]{\endgroup#1\@@endlink}%
\providecommand \@sanitize@url [0]{\catcode `\\12\catcode `\$12\catcode
  `\&12\catcode `\#12\catcode `\^12\catcode `\_12\catcode `\%12\relax}%
\providecommand \@@startlink[1]{}%
\providecommand \@@endlink[0]{}%
\providecommand \url  [0]{\begingroup\@sanitize@url \@url }%
\providecommand \@url [1]{\endgroup\@href {#1}{\urlprefix }}%
\providecommand \urlprefix  [0]{URL }%
\providecommand \Eprint [0]{\href }%
\providecommand \doibase [0]{http://dx.doi.org/}%
\providecommand \selectlanguage [0]{\@gobble}%
\providecommand \bibinfo  [0]{\@secondoftwo}%
\providecommand \bibfield  [0]{\@secondoftwo}%
\providecommand \translation [1]{[#1]}%
\providecommand \BibitemOpen [0]{}%
\providecommand \bibitemStop [0]{}%
\providecommand \bibitemNoStop [0]{.\EOS\space}%
\providecommand \EOS [0]{\spacefactor3000\relax}%
\providecommand \BibitemShut  [1]{\csname bibitem#1\endcsname}%
\let\auto@bib@innerbib\@empty
\bibitem [{\citenamefont {Schr{\"o}dinger}(1935)}]{Schrodinger:1935}%
  \BibitemOpen
  \bibfield  {author} {\bibinfo {author} {\bibfnamefont {E.}~\bibnamefont
  {Schr{\"o}dinger}},\ }\href@noop {} {\bibfield  {journal} {\bibinfo
  {journal} {Naturwiss.}\ }\textbf {\bibinfo {volume} {23}},\ \bibinfo {pages}
  {807} (\bibinfo {year} {1935})}\BibitemShut {NoStop}%
\bibitem [{\citenamefont {Einstein}\ \emph {et~al.}(1935)\citenamefont
  {Einstein}, \citenamefont {Podolsky},\ and\ \citenamefont
  {Rosen}}]{Einstein:1935}%
  \BibitemOpen
  \bibfield  {author} {\bibinfo {author} {\bibfnamefont {A.}~\bibnamefont
  {Einstein}}, \bibinfo {author} {\bibfnamefont {B.}~\bibnamefont {Podolsky}},
  \ and\ \bibinfo {author} {\bibfnamefont {N.}~\bibnamefont {Rosen}},\
  }\href@noop {} {\bibfield  {journal} {\bibinfo  {journal} {Phys. Rev}\
  }\textbf {\bibinfo {volume} {47}},\ \bibinfo {pages} {777} (\bibinfo {year}
  {1935})}\BibitemShut {NoStop}%
\bibitem [{\citenamefont {Horodecki}\ \emph {et~al.}(2009)\citenamefont
  {Horodecki}, \citenamefont {Horodecki}, \citenamefont {Horodecki},\ and\
  \citenamefont {Horodecki}}]{Horodecki:2009}%
  \BibitemOpen
  \bibfield  {author} {\bibinfo {author} {\bibfnamefont {R.}~\bibnamefont
  {Horodecki}}, \bibinfo {author} {\bibfnamefont {P.}~\bibnamefont
  {Horodecki}}, \bibinfo {author} {\bibfnamefont {M.}~\bibnamefont
  {Horodecki}}, \ and\ \bibinfo {author} {\bibfnamefont {K.}~\bibnamefont
  {Horodecki}},\ }\href@noop {} {\bibfield  {journal} {\bibinfo  {journal}
  {Rev. Mod. Phys.}\ }\textbf {\bibinfo {volume} {81}},\ \bibinfo {pages} {865}
  (\bibinfo {year} {2009})}\BibitemShut {NoStop}%
\bibitem [{\citenamefont {G{\"u}hne}\ and\ \citenamefont
  {T{\'o}th}(2009)}]{Guhne:2009}%
  \BibitemOpen
  \bibfield  {author} {\bibinfo {author} {\bibfnamefont {O.}~\bibnamefont
  {G{\"u}hne}}\ and\ \bibinfo {author} {\bibfnamefont {G.}~\bibnamefont
  {T{\'o}th}},\ }\href@noop {} {\bibfield  {journal} {\bibinfo  {journal}
  {Phys. Rep.}\ }\textbf {\bibinfo {volume} {474}},\ \bibinfo {pages} {1 }
  (\bibinfo {year} {2009})}\BibitemShut {NoStop}%
\bibitem [{\citenamefont {Plenio}\ and\ \citenamefont
  {Virmani}(2007)}]{Plenio:2007}%
  \BibitemOpen
  \bibfield  {author} {\bibinfo {author} {\bibfnamefont {M.~B.}\ \bibnamefont
  {Plenio}}\ and\ \bibinfo {author} {\bibfnamefont {S.}~\bibnamefont
  {Virmani}},\ }\href@noop {} {\bibfield  {journal} {\bibinfo  {journal}
  {Quant. Inf. Comp.}\ }\textbf {\bibinfo {volume} {7}},\ \bibinfo {pages} {1}
  (\bibinfo {year} {2007})}\BibitemShut {NoStop}%
\bibitem [{\citenamefont {Bennett}\ \emph {et~al.}(1996)\citenamefont
  {Bennett}, \citenamefont {Brassard}, \citenamefont {Popescu}, \citenamefont
  {Schumacher}, \citenamefont {Smolin},\ and\ \citenamefont
  {Wootters}}]{Bennett:1996}%
  \BibitemOpen
  \bibfield  {author} {\bibinfo {author} {\bibfnamefont {C.~H.}\ \bibnamefont
  {Bennett}}, \bibinfo {author} {\bibfnamefont {G.}~\bibnamefont {Brassard}},
  \bibinfo {author} {\bibfnamefont {S.}~\bibnamefont {Popescu}}, \bibinfo
  {author} {\bibfnamefont {B.}~\bibnamefont {Schumacher}}, \bibinfo {author}
  {\bibfnamefont {J.~A.}\ \bibnamefont {Smolin}}, \ and\ \bibinfo {author}
  {\bibfnamefont {W.~K.}\ \bibnamefont {Wootters}},\ }\href@noop {} {\bibfield
  {journal} {\bibinfo  {journal} {Phys. Rev. Lett.}\ }\textbf {\bibinfo
  {volume} {76}},\ \bibinfo {pages} {722} (\bibinfo {year} {1996})}\BibitemShut
  {NoStop}%
\bibitem [{\citenamefont {Giovannetti}\ \emph {et~al.}(2011)\citenamefont
  {Giovannetti}, \citenamefont {Lloyd},\ and\ \citenamefont
  {Maccone}}]{Giovannetti:2011}%
  \BibitemOpen
  \bibfield  {author} {\bibinfo {author} {\bibfnamefont {V.}~\bibnamefont
  {Giovannetti}}, \bibinfo {author} {\bibfnamefont {S.}~\bibnamefont {Lloyd}},
  \ and\ \bibinfo {author} {\bibfnamefont {L.}~\bibnamefont {Maccone}},\
  }\href@noop {} {\bibfield  {journal} {\bibinfo  {journal} {Nat. Phot.}\
  }\textbf {\bibinfo {volume} {5}},\ \bibinfo {pages} {222} (\bibinfo {year}
  {2011})}\BibitemShut {NoStop}%
\bibitem [{\citenamefont {Gisin}\ \emph {et~al.}(2002)\citenamefont {Gisin},
  \citenamefont {Ribordy}, \citenamefont {Tittel},\ and\ \citenamefont
  {Zbinden}}]{Gisin:2002}%
  \BibitemOpen
  \bibfield  {author} {\bibinfo {author} {\bibfnamefont {N.}~\bibnamefont
  {Gisin}}, \bibinfo {author} {\bibfnamefont {G.}~\bibnamefont {Ribordy}},
  \bibinfo {author} {\bibfnamefont {W.}~\bibnamefont {Tittel}}, \ and\ \bibinfo
  {author} {\bibfnamefont {H.}~\bibnamefont {Zbinden}},\ }\href@noop {}
  {\bibfield  {journal} {\bibinfo  {journal} {Rev. Mod. Phys.}\ }\textbf
  {\bibinfo {volume} {74}},\ \bibinfo {pages} {145} (\bibinfo {year}
  {2002})}\BibitemShut {NoStop}%
\bibitem [{\citenamefont {Nielsen}\ and\ \citenamefont
  {Chuang}(2000)}]{Nielsen:2000}%
  \BibitemOpen
  \bibfield  {author} {\bibinfo {author} {\bibfnamefont {M.~A.}\ \bibnamefont
  {Nielsen}}\ and\ \bibinfo {author} {\bibfnamefont {I.~L.}\ \bibnamefont
  {Chuang}},\ }\href@noop {} {\emph {\bibinfo {title} {Quantum Information}}}\
  (\bibinfo  {publisher} {Cambridge University Press},\ \bibinfo {year}
  {2000})\BibitemShut {NoStop}%
\bibitem [{\citenamefont {Preskill}(2000)}]{Preskill:2000}%
  \BibitemOpen
  \bibfield  {author} {\bibinfo {author} {\bibfnamefont {J.}~\bibnamefont
  {Preskill}},\ }\href@noop {} {\bibfield  {journal} {\bibinfo  {journal} {J.
  Mod. Opt.}\ }\textbf {\bibinfo {volume} {47}},\ \bibinfo {pages} {127}
  (\bibinfo {year} {2000})}\BibitemShut {NoStop}%
\bibitem [{\citenamefont {Amico}\ \emph {et~al.}(2008)\citenamefont {Amico},
  \citenamefont {Fazio}, \citenamefont {Osterloh},\ and\ \citenamefont
  {Vedral}}]{Amico:2008}%
  \BibitemOpen
  \bibfield  {author} {\bibinfo {author} {\bibfnamefont {L.}~\bibnamefont
  {Amico}}, \bibinfo {author} {\bibfnamefont {R.}~\bibnamefont {Fazio}},
  \bibinfo {author} {\bibfnamefont {A.}~\bibnamefont {Osterloh}}, \ and\
  \bibinfo {author} {\bibfnamefont {V.}~\bibnamefont {Vedral}},\ }\href@noop {}
  {\bibfield  {journal} {\bibinfo  {journal} {Rev. Mod. Phys.}\ }\textbf
  {\bibinfo {volume} {80}},\ \bibinfo {pages} {517} (\bibinfo {year}
  {2008})}\BibitemShut {NoStop}%
\bibitem [{\citenamefont {De~Chiara}\ and\ \citenamefont
  {Sanpera}(2018)}]{DeChiara:2018}%
  \BibitemOpen
  \bibfield  {author} {\bibinfo {author} {\bibfnamefont {G.}~\bibnamefont
  {De~Chiara}}\ and\ \bibinfo {author} {\bibfnamefont {A.}~\bibnamefont
  {Sanpera}},\ }\href@noop {} {\bibfield  {journal} {\bibinfo  {journal} {Rep.
  Prog. Phys.}\ }\textbf {\bibinfo {volume} {81}},\ \bibinfo {pages} {074002}
  (\bibinfo {year} {2018})}\BibitemShut {NoStop}%
\bibitem [{\citenamefont {Islam}\ \emph {et~al.}(2015)\citenamefont {Islam},
  \citenamefont {Ma}, \citenamefont {Preiss}, \citenamefont {Tai},
  \citenamefont {Lukin}, \citenamefont {Rispoli},\ and\ \citenamefont
  {Greiner}}]{Islam:2015}%
  \BibitemOpen
  \bibfield  {author} {\bibinfo {author} {\bibfnamefont {R.}~\bibnamefont
  {Islam}}, \bibinfo {author} {\bibfnamefont {R.}~\bibnamefont {Ma}}, \bibinfo
  {author} {\bibfnamefont {P.~M.}\ \bibnamefont {Preiss}}, \bibinfo {author}
  {\bibfnamefont {M.~E.}\ \bibnamefont {Tai}}, \bibinfo {author} {\bibfnamefont
  {A.}~\bibnamefont {Lukin}}, \bibinfo {author} {\bibfnamefont
  {M.}~\bibnamefont {Rispoli}}, \ and\ \bibinfo {author} {\bibfnamefont
  {M.}~\bibnamefont {Greiner}},\ }\href@noop {} {\bibfield  {journal} {\bibinfo
   {journal} {Nature}\ }\textbf {\bibinfo {volume} {528}},\ \bibinfo {pages}
  {77} (\bibinfo {year} {2015})}\BibitemShut {NoStop}%
\bibitem [{\citenamefont {Amico}\ \emph {et~al.}(2004)\citenamefont {Amico},
  \citenamefont {Osterloh}, \citenamefont {Plastina}, \citenamefont {Fazio},\
  and\ \citenamefont {Palma}}]{amicospins}%
  \BibitemOpen
  \bibfield  {author} {\bibinfo {author} {\bibfnamefont {L.}~\bibnamefont
  {Amico}}, \bibinfo {author} {\bibfnamefont {A.}~\bibnamefont {Osterloh}},
  \bibinfo {author} {\bibfnamefont {F.}~\bibnamefont {Plastina}}, \bibinfo
  {author} {\bibfnamefont {R.}~\bibnamefont {Fazio}}, \ and\ \bibinfo {author}
  {\bibfnamefont {G.~M.}\ \bibnamefont {Palma}},\ }\href@noop {} {\bibfield
  {journal} {\bibinfo  {journal} {Phys. Rev. A}\ }\textbf {\bibinfo {volume}
  {69}},\ \bibinfo {pages} {022304} (\bibinfo {year} {2004})}\BibitemShut
  {NoStop}%
\bibitem [{\citenamefont {Jurcevic}\ \emph {et~al.}(2014)\citenamefont
  {Jurcevic}, \citenamefont {Lanyon}, \citenamefont {Hauke}, \citenamefont
  {Hempel}, \citenamefont {Zoller}, \citenamefont {Blatt},\ and\ \citenamefont
  {Roos}}]{Jurcevic:2014}%
  \BibitemOpen
  \bibfield  {author} {\bibinfo {author} {\bibfnamefont {P.}~\bibnamefont
  {Jurcevic}}, \bibinfo {author} {\bibfnamefont {B.~P.}\ \bibnamefont
  {Lanyon}}, \bibinfo {author} {\bibfnamefont {P.}~\bibnamefont {Hauke}},
  \bibinfo {author} {\bibfnamefont {C.}~\bibnamefont {Hempel}}, \bibinfo
  {author} {\bibfnamefont {P.}~\bibnamefont {Zoller}}, \bibinfo {author}
  {\bibfnamefont {R.}~\bibnamefont {Blatt}}, \ and\ \bibinfo {author}
  {\bibfnamefont {C.~F.}\ \bibnamefont {Roos}},\ }\href@noop {} {\bibfield
  {journal} {\bibinfo  {journal} {Nature}\ }\textbf {\bibinfo {volume} {511}},\
  \bibinfo {pages} {202} (\bibinfo {year} {2014})}\BibitemShut {NoStop}%
\bibitem [{\citenamefont {Kastner}(2015)}]{Kastner:2015}%
  \BibitemOpen
  \bibfield  {author} {\bibinfo {author} {\bibfnamefont {M.}~\bibnamefont
  {Kastner}},\ }\href@noop {} {\bibfield  {journal} {\bibinfo  {journal} {New
  J. Phys.}\ }\textbf {\bibinfo {volume} {17}},\ \bibinfo {pages} {123024}
  (\bibinfo {year} {2015})}\BibitemShut {NoStop}%
\bibitem [{\citenamefont {Alba}\ and\ \citenamefont
  {Carollo}(2020)}]{alba2020spreading}%
  \BibitemOpen
  \bibfield  {author} {\bibinfo {author} {\bibfnamefont {V.}~\bibnamefont
  {Alba}}\ and\ \bibinfo {author} {\bibfnamefont {F.}~\bibnamefont {Carollo}},\
  }\href@noop {} {\bibfield  {journal} {\bibinfo  {journal} {arXiv e-prints,
  2002.09527, cond-mat.stat-mech}\ } (\bibinfo {year} {2020})}\BibitemShut
  {NoStop}%
\bibitem [{\citenamefont {L{\"a}uchli}\ and\ \citenamefont
  {Kollath}(2008)}]{Lauchli:2008}%
  \BibitemOpen
  \bibfield  {author} {\bibinfo {author} {\bibfnamefont {A.~M.}\ \bibnamefont
  {L{\"a}uchli}}\ and\ \bibinfo {author} {\bibfnamefont {C.}~\bibnamefont
  {Kollath}},\ }\href@noop {} {\bibfield  {journal} {\bibinfo  {journal} {J.
  Stat. Mech.}\ }\textbf {\bibinfo {volume} {2008}},\ \bibinfo {pages} {P05018}
  (\bibinfo {year} {2008})}\BibitemShut {NoStop}%
\bibitem [{\citenamefont {Cheneau}\ \emph {et~al.}(2012)\citenamefont
  {Cheneau}, \citenamefont {Barmettler}, \citenamefont {Poletti}, \citenamefont
  {Endres}, \citenamefont {Schauß}, \citenamefont {Fukuhara}, \citenamefont
  {Gross}, \citenamefont {Bloch}, \citenamefont {Kollath},\ and\ \citenamefont
  {Kuhr}}]{cheneau_light-cone-like_2012}%
  \BibitemOpen
  \bibfield  {author} {\bibinfo {author} {\bibfnamefont {M.}~\bibnamefont
  {Cheneau}}, \bibinfo {author} {\bibfnamefont {P.}~\bibnamefont {Barmettler}},
  \bibinfo {author} {\bibfnamefont {D.}~\bibnamefont {Poletti}}, \bibinfo
  {author} {\bibfnamefont {M.}~\bibnamefont {Endres}}, \bibinfo {author}
  {\bibfnamefont {P.}~\bibnamefont {Schauß}}, \bibinfo {author} {\bibfnamefont
  {T.}~\bibnamefont {Fukuhara}}, \bibinfo {author} {\bibfnamefont
  {C.}~\bibnamefont {Gross}}, \bibinfo {author} {\bibfnamefont
  {I.}~\bibnamefont {Bloch}}, \bibinfo {author} {\bibfnamefont
  {C.}~\bibnamefont {Kollath}}, \ and\ \bibinfo {author} {\bibfnamefont
  {S.}~\bibnamefont {Kuhr}},\ }\href@noop {} {\bibfield  {journal} {\bibinfo
  {journal} {Nature}\ }\textbf {\bibinfo {volume} {481}},\ \bibinfo {pages}
  {484} (\bibinfo {year} {2012})}\BibitemShut {NoStop}%
\bibitem [{\citenamefont {Daley}\ \emph {et~al.}(2012)\citenamefont {Daley},
  \citenamefont {Pichler}, \citenamefont {Schachenmayer},\ and\ \citenamefont
  {Zoller}}]{Daley:2012}%
  \BibitemOpen
  \bibfield  {author} {\bibinfo {author} {\bibfnamefont {A.~J.}\ \bibnamefont
  {Daley}}, \bibinfo {author} {\bibfnamefont {H.}~\bibnamefont {Pichler}},
  \bibinfo {author} {\bibfnamefont {J.}~\bibnamefont {Schachenmayer}}, \ and\
  \bibinfo {author} {\bibfnamefont {P.}~\bibnamefont {Zoller}},\ }\href@noop {}
  {\bibfield  {journal} {\bibinfo  {journal} {Phys. Rev. Lett.}\ }\textbf
  {\bibinfo {volume} {109}},\ \bibinfo {pages} {020505} (\bibinfo {year}
  {2012})}\BibitemShut {NoStop}%
\bibitem [{\citenamefont {Bernier}\ \emph {et~al.}(2018)\citenamefont
  {Bernier}, \citenamefont {Tan}, \citenamefont {Bonnes}, \citenamefont {Guo},
  \citenamefont {Poletti},\ and\ \citenamefont
  {Kollath}}]{bernier_light-cone_2018}%
  \BibitemOpen
  \bibfield  {author} {\bibinfo {author} {\bibfnamefont {J.-S.}\ \bibnamefont
  {Bernier}}, \bibinfo {author} {\bibfnamefont {R.}~\bibnamefont {Tan}},
  \bibinfo {author} {\bibfnamefont {L.}~\bibnamefont {Bonnes}}, \bibinfo
  {author} {\bibfnamefont {C.}~\bibnamefont {Guo}}, \bibinfo {author}
  {\bibfnamefont {D.}~\bibnamefont {Poletti}}, \ and\ \bibinfo {author}
  {\bibfnamefont {C.}~\bibnamefont {Kollath}},\ }\href@noop {} {\bibfield
  {journal} {\bibinfo  {journal} {Phys. Rev. Lett.}\ }\textbf {\bibinfo
  {volume} {120}},\ \bibinfo {pages} {020401} (\bibinfo {year}
  {2018})}\BibitemShut {NoStop}%
\bibitem [{\citenamefont {Aolita}\ \emph {et~al.}(2015)\citenamefont {Aolita},
  \citenamefont {de~Melo},\ and\ \citenamefont {Davidovich}}]{Aolita_2015}%
  \BibitemOpen
  \bibfield  {author} {\bibinfo {author} {\bibfnamefont {L.}~\bibnamefont
  {Aolita}}, \bibinfo {author} {\bibfnamefont {F.}~\bibnamefont {de~Melo}}, \
  and\ \bibinfo {author} {\bibfnamefont {L.}~\bibnamefont {Davidovich}},\
  }\href@noop {} {\bibfield  {journal} {\bibinfo  {journal} {Reports on
  Progress in Physics}\ }\textbf {\bibinfo {volume} {78}},\ \bibinfo {pages}
  {042001} (\bibinfo {year} {2015})}\BibitemShut {NoStop}%
\bibitem [{\citenamefont {Leghtas}\ \emph {et~al.}(2015)\citenamefont
  {Leghtas}, \citenamefont {Touzard}, \citenamefont {Pop}, \citenamefont {Kou},
  \citenamefont {Vlastakis}, \citenamefont {Petrenko}, \citenamefont {Sliwa},
  \citenamefont {Narla}, \citenamefont {Shankar}, \citenamefont {Hatridge},
  \citenamefont {Reagor}, \citenamefont {Frunzio}, \citenamefont {Schoelkopf},
  \citenamefont {Mirrahimi},\ and\ \citenamefont {Devoret}}]{Leghtas853}%
  \BibitemOpen
  \bibfield  {author} {\bibinfo {author} {\bibfnamefont {Z.}~\bibnamefont
  {Leghtas}}, \bibinfo {author} {\bibfnamefont {S.}~\bibnamefont {Touzard}},
  \bibinfo {author} {\bibfnamefont {I.~M.}\ \bibnamefont {Pop}}, \bibinfo
  {author} {\bibfnamefont {A.}~\bibnamefont {Kou}}, \bibinfo {author}
  {\bibfnamefont {B.}~\bibnamefont {Vlastakis}}, \bibinfo {author}
  {\bibfnamefont {A.}~\bibnamefont {Petrenko}}, \bibinfo {author}
  {\bibfnamefont {K.~M.}\ \bibnamefont {Sliwa}}, \bibinfo {author}
  {\bibfnamefont {A.}~\bibnamefont {Narla}}, \bibinfo {author} {\bibfnamefont
  {S.}~\bibnamefont {Shankar}}, \bibinfo {author} {\bibfnamefont {M.~J.}\
  \bibnamefont {Hatridge}}, \bibinfo {author} {\bibfnamefont {M.}~\bibnamefont
  {Reagor}}, \bibinfo {author} {\bibfnamefont {L.}~\bibnamefont {Frunzio}},
  \bibinfo {author} {\bibfnamefont {R.~J.}\ \bibnamefont {Schoelkopf}},
  \bibinfo {author} {\bibfnamefont {M.}~\bibnamefont {Mirrahimi}}, \ and\
  \bibinfo {author} {\bibfnamefont {M.~H.}\ \bibnamefont {Devoret}},\
  }\href@noop {} {\bibfield  {journal} {\bibinfo  {journal} {Science}\ }\textbf
  {\bibinfo {volume} {347}},\ \bibinfo {pages} {853} (\bibinfo {year}
  {2015})}\BibitemShut {NoStop}%
\bibitem [{\citenamefont {Ma}\ \emph {et~al.}(2019)\citenamefont {Ma},
  \citenamefont {Saxberg}, \citenamefont {Owens}, \citenamefont {Leung},
  \citenamefont {Lu}, \citenamefont {Simon},\ and\ \citenamefont
  {Schuster}}]{Mott}%
  \BibitemOpen
  \bibfield  {author} {\bibinfo {author} {\bibfnamefont {R.}~\bibnamefont
  {Ma}}, \bibinfo {author} {\bibfnamefont {B.}~\bibnamefont {Saxberg}},
  \bibinfo {author} {\bibfnamefont {C.}~\bibnamefont {Owens}}, \bibinfo
  {author} {\bibfnamefont {N.}~\bibnamefont {Leung}}, \bibinfo {author}
  {\bibfnamefont {Y.}~\bibnamefont {Lu}}, \bibinfo {author} {\bibfnamefont
  {J.}~\bibnamefont {Simon}}, \ and\ \bibinfo {author} {\bibfnamefont {D.~I.}\
  \bibnamefont {Schuster}},\ }\href@noop {} {\bibfield  {journal} {\bibinfo
  {journal} {Nature}\ }\textbf {\bibinfo {volume} {566}},\ \bibinfo {pages}
  {51} (\bibinfo {year} {2019})}\BibitemShut {NoStop}%
\bibitem [{\citenamefont {Carusotto}\ and\ \citenamefont
  {Ciuti}(2013)}]{rmpciuti}%
  \BibitemOpen
  \bibfield  {author} {\bibinfo {author} {\bibfnamefont {I.}~\bibnamefont
  {Carusotto}}\ and\ \bibinfo {author} {\bibfnamefont {C.}~\bibnamefont
  {Ciuti}},\ }\href@noop {} {\bibfield  {journal} {\bibinfo  {journal} {Rev.
  Mod. Phys.}\ }\textbf {\bibinfo {volume} {85}},\ \bibinfo {pages} {299}
  (\bibinfo {year} {2013})}\BibitemShut {NoStop}%
\bibitem [{\citenamefont {Schmidt}\ and\ \citenamefont
  {Koch}(2013)}]{Schmidt:2013}%
  \BibitemOpen
  \bibfield  {author} {\bibinfo {author} {\bibfnamefont {S.}~\bibnamefont
  {Schmidt}}\ and\ \bibinfo {author} {\bibfnamefont {J.}~\bibnamefont {Koch}},\
  }\href@noop {} {\bibfield  {journal} {\bibinfo  {journal} {Ann. Phys.}\
  }\textbf {\bibinfo {volume} {525}},\ \bibinfo {pages} {395} (\bibinfo {year}
  {2013})}\BibitemShut {NoStop}%
\bibitem [{\citenamefont {Hartmann}(2016{\natexlab{a}})}]{HartmannRev2016}%
  \BibitemOpen
  \bibfield  {author} {\bibinfo {author} {\bibfnamefont {M.~J.}\ \bibnamefont
  {Hartmann}},\ }\href@noop {} {\bibfield  {journal} {\bibinfo  {journal}
  {Journal of Optics}\ }\textbf {\bibinfo {volume} {18}},\ \bibinfo {pages}
  {104005} (\bibinfo {year} {2016}{\natexlab{a}})}\BibitemShut {NoStop}%
\bibitem [{\citenamefont {Noh}\ and\ \citenamefont
  {Angelakis}(2016)}]{AngelakisRev2016}%
  \BibitemOpen
  \bibfield  {author} {\bibinfo {author} {\bibfnamefont {C.}~\bibnamefont
  {Noh}}\ and\ \bibinfo {author} {\bibfnamefont {D.~G.}\ \bibnamefont
  {Angelakis}},\ }\href@noop {} {\bibfield  {journal} {\bibinfo  {journal}
  {Reports on Progress in Physics}\ }\textbf {\bibinfo {volume} {80}},\
  \bibinfo {pages} {016401} (\bibinfo {year} {2016})}\BibitemShut {NoStop}%
\bibitem [{\citenamefont {Roy}\ \emph {et~al.}(2017)\citenamefont {Roy},
  \citenamefont {Wilson},\ and\ \citenamefont {Firstenberg}}]{Roy:2017}%
  \BibitemOpen
  \bibfield  {author} {\bibinfo {author} {\bibfnamefont {D.}~\bibnamefont
  {Roy}}, \bibinfo {author} {\bibfnamefont {C.~M.}\ \bibnamefont {Wilson}}, \
  and\ \bibinfo {author} {\bibfnamefont {O.}~\bibnamefont {Firstenberg}},\
  }\href@noop {} {\bibfield  {journal} {\bibinfo  {journal} {Rev. Mod. Phys.}\
  }\textbf {\bibinfo {volume} {89}},\ \bibinfo {pages} {021001} (\bibinfo
  {year} {2017})}\BibitemShut {NoStop}%
\bibitem [{\citenamefont {Biella}\ \emph {et~al.}(2015)\citenamefont {Biella},
  \citenamefont {Mazza}, \citenamefont {Carusotto}, \citenamefont {Rossini},\
  and\ \citenamefont {Fazio}}]{Biella:2015}%
  \BibitemOpen
  \bibfield  {author} {\bibinfo {author} {\bibfnamefont {A.}~\bibnamefont
  {Biella}}, \bibinfo {author} {\bibfnamefont {L.}~\bibnamefont {Mazza}},
  \bibinfo {author} {\bibfnamefont {I.}~\bibnamefont {Carusotto}}, \bibinfo
  {author} {\bibfnamefont {D.}~\bibnamefont {Rossini}}, \ and\ \bibinfo
  {author} {\bibfnamefont {R.}~\bibnamefont {Fazio}},\ }\href@noop {}
  {\bibfield  {journal} {\bibinfo  {journal} {Phys. Rev. A}\ }\textbf {\bibinfo
  {volume} {91}},\ \bibinfo {pages} {053815} (\bibinfo {year}
  {2015})}\BibitemShut {NoStop}%
\bibitem [{\citenamefont {Lee}\ \emph {et~al.}(2015)\citenamefont {Lee},
  \citenamefont {Noh}, \citenamefont {Schetakis},\ and\ \citenamefont
  {Angelakis}}]{Lee:2015}%
  \BibitemOpen
  \bibfield  {author} {\bibinfo {author} {\bibfnamefont {C.}~\bibnamefont
  {Lee}}, \bibinfo {author} {\bibfnamefont {C.}~\bibnamefont {Noh}}, \bibinfo
  {author} {\bibfnamefont {N.}~\bibnamefont {Schetakis}}, \ and\ \bibinfo
  {author} {\bibfnamefont {D.~G.}\ \bibnamefont {Angelakis}},\ }\href@noop {}
  {\bibfield  {journal} {\bibinfo  {journal} {Phys. Rev. A}\ }\textbf {\bibinfo
  {volume} {92}},\ \bibinfo {pages} {063817} (\bibinfo {year}
  {2015})}\BibitemShut {NoStop}%
\bibitem [{\citenamefont {Mertz}\ \emph {et~al.}(2016)\citenamefont {Mertz},
  \citenamefont {Vasi{\'c}}, \citenamefont {Hartmann},\ and\ \citenamefont
  {Hofstetter}}]{Mertz:2016}%
  \BibitemOpen
  \bibfield  {author} {\bibinfo {author} {\bibfnamefont {T.}~\bibnamefont
  {Mertz}}, \bibinfo {author} {\bibfnamefont {I.}~\bibnamefont {Vasi{\'c}}},
  \bibinfo {author} {\bibfnamefont {M.~J.}\ \bibnamefont {Hartmann}}, \ and\
  \bibinfo {author} {\bibfnamefont {W.}~\bibnamefont {Hofstetter}},\
  }\href@noop {} {\bibfield  {journal} {\bibinfo  {journal} {Physical Review
  A}\ }\textbf {\bibinfo {volume} {94}},\ \bibinfo {pages} {013809} (\bibinfo
  {year} {2016})}\BibitemShut {NoStop}%
\bibitem [{\citenamefont {Debnath}\ \emph {et~al.}(2017)\citenamefont
  {Debnath}, \citenamefont {Mascarenhas},\ and\ \citenamefont
  {Savona}}]{Debnath:2017}%
  \BibitemOpen
  \bibfield  {author} {\bibinfo {author} {\bibfnamefont {K.}~\bibnamefont
  {Debnath}}, \bibinfo {author} {\bibfnamefont {E.}~\bibnamefont
  {Mascarenhas}}, \ and\ \bibinfo {author} {\bibfnamefont {V.}~\bibnamefont
  {Savona}},\ }\href@noop {} {\bibfield  {journal} {\bibinfo  {journal} {New
  Journal of Physics}\ }\textbf {\bibinfo {volume} {19}},\ \bibinfo {pages}
  {115006} (\bibinfo {year} {2017})}\BibitemShut {NoStop}%
\bibitem [{\citenamefont {Houck}\ \emph {et~al.}(2012)\citenamefont {Houck},
  \citenamefont {T{\"u}reci},\ and\ \citenamefont {Koch}}]{Houck2012}%
  \BibitemOpen
  \bibfield  {author} {\bibinfo {author} {\bibfnamefont {A.~A.}\ \bibnamefont
  {Houck}}, \bibinfo {author} {\bibfnamefont {H.~E.}\ \bibnamefont
  {T{\"u}reci}}, \ and\ \bibinfo {author} {\bibfnamefont {J.}~\bibnamefont
  {Koch}},\ }\href@noop {} {\bibfield  {journal} {\bibinfo  {journal} {Nature
  Physics}\ }\textbf {\bibinfo {volume} {8}},\ \bibinfo {pages} {292} (\bibinfo
  {year} {2012})}\BibitemShut {NoStop}%
\bibitem [{\citenamefont {Fitzpatrick}\ \emph {et~al.}(2017)\citenamefont
  {Fitzpatrick}, \citenamefont {Sundaresan}, \citenamefont {Li}, \citenamefont
  {Koch},\ and\ \citenamefont {Houck}}]{cQEDTransition}%
  \BibitemOpen
  \bibfield  {author} {\bibinfo {author} {\bibfnamefont {M.}~\bibnamefont
  {Fitzpatrick}}, \bibinfo {author} {\bibfnamefont {N.~M.}\ \bibnamefont
  {Sundaresan}}, \bibinfo {author} {\bibfnamefont {A.~C.~Y.}\ \bibnamefont
  {Li}}, \bibinfo {author} {\bibfnamefont {J.}~\bibnamefont {Koch}}, \ and\
  \bibinfo {author} {\bibfnamefont {A.~A.}\ \bibnamefont {Houck}},\ }\href@noop
  {} {\bibfield  {journal} {\bibinfo  {journal} {Phys. Rev. X}\ }\textbf
  {\bibinfo {volume} {7}},\ \bibinfo {pages} {011016} (\bibinfo {year}
  {2017})}\BibitemShut {NoStop}%
\bibitem [{\citenamefont {Carusotto}\ \emph {et~al.}(2020)\citenamefont
  {Carusotto}, \citenamefont {Houck}, \citenamefont {Koll{\'a}r}, \citenamefont
  {Roushan}, \citenamefont {Schuster},\ and\ \citenamefont
  {Simon}}]{Carusotto2020}%
  \BibitemOpen
  \bibfield  {author} {\bibinfo {author} {\bibfnamefont {I.}~\bibnamefont
  {Carusotto}}, \bibinfo {author} {\bibfnamefont {A.~A.}\ \bibnamefont
  {Houck}}, \bibinfo {author} {\bibfnamefont {A.~J.}\ \bibnamefont
  {Koll{\'a}r}}, \bibinfo {author} {\bibfnamefont {P.}~\bibnamefont {Roushan}},
  \bibinfo {author} {\bibfnamefont {D.~I.}\ \bibnamefont {Schuster}}, \ and\
  \bibinfo {author} {\bibfnamefont {J.}~\bibnamefont {Simon}},\ }\href@noop {}
  {\bibfield  {journal} {\bibinfo  {journal} {Nature Physics}\ }\textbf
  {\bibinfo {volume} {16}},\ \bibinfo {pages} {268} (\bibinfo {year}
  {2020})}\BibitemShut {NoStop}%
\bibitem [{\citenamefont {Goblot}\ \emph {et~al.}(2019)\citenamefont {Goblot},
  \citenamefont {Rauer}, \citenamefont {Vicentini}, \citenamefont {Le~Boit\'e},
  \citenamefont {Galopin}, \citenamefont {Lema\^{\i}tre}, \citenamefont
  {Le~Gratiet}, \citenamefont {Harouri}, \citenamefont {Sagnes}, \citenamefont
  {Ravets}, \citenamefont {Ciuti}, \citenamefont {Amo},\ and\ \citenamefont
  {Bloch}}]{bloch}%
  \BibitemOpen
  \bibfield  {author} {\bibinfo {author} {\bibfnamefont {V.}~\bibnamefont
  {Goblot}}, \bibinfo {author} {\bibfnamefont {B.}~\bibnamefont {Rauer}},
  \bibinfo {author} {\bibfnamefont {F.}~\bibnamefont {Vicentini}}, \bibinfo
  {author} {\bibfnamefont {A.}~\bibnamefont {Le~Boit\'e}}, \bibinfo {author}
  {\bibfnamefont {E.}~\bibnamefont {Galopin}}, \bibinfo {author} {\bibfnamefont
  {A.}~\bibnamefont {Lema\^{\i}tre}}, \bibinfo {author} {\bibfnamefont
  {L.}~\bibnamefont {Le~Gratiet}}, \bibinfo {author} {\bibfnamefont
  {A.}~\bibnamefont {Harouri}}, \bibinfo {author} {\bibfnamefont
  {I.}~\bibnamefont {Sagnes}}, \bibinfo {author} {\bibfnamefont
  {S.}~\bibnamefont {Ravets}}, \bibinfo {author} {\bibfnamefont
  {C.}~\bibnamefont {Ciuti}}, \bibinfo {author} {\bibfnamefont
  {A.}~\bibnamefont {Amo}}, \ and\ \bibinfo {author} {\bibfnamefont
  {J.}~\bibnamefont {Bloch}},\ }\href@noop {} {\bibfield  {journal} {\bibinfo
  {journal} {Phys. Rev. Lett.}\ }\textbf {\bibinfo {volume} {123}},\ \bibinfo
  {pages} {113901} (\bibinfo {year} {2019})}\BibitemShut {NoStop}%
\bibitem [{\citenamefont {Greiner}\ \emph {et~al.}(2002)\citenamefont
  {Greiner}, \citenamefont {Mandel}, \citenamefont {Esslinger}, \citenamefont
  {H{\"a}nsch},\ and\ \citenamefont {Bloch}}]{Greiner2002}%
  \BibitemOpen
  \bibfield  {author} {\bibinfo {author} {\bibfnamefont {M.}~\bibnamefont
  {Greiner}}, \bibinfo {author} {\bibfnamefont {O.}~\bibnamefont {Mandel}},
  \bibinfo {author} {\bibfnamefont {T.}~\bibnamefont {Esslinger}}, \bibinfo
  {author} {\bibfnamefont {T.~W.}\ \bibnamefont {H{\"a}nsch}}, \ and\ \bibinfo
  {author} {\bibfnamefont {I.}~\bibnamefont {Bloch}},\ }\href@noop {}
  {\bibfield  {journal} {\bibinfo  {journal} {Nature}\ }\textbf {\bibinfo
  {volume} {415}},\ \bibinfo {pages} {39} (\bibinfo {year} {2002})}\BibitemShut
  {NoStop}%
\bibitem [{\citenamefont {Jaksch}\ \emph {et~al.}(1998)\citenamefont {Jaksch},
  \citenamefont {Bruder}, \citenamefont {Cirac}, \citenamefont {Gardiner},\
  and\ \citenamefont {Zoller}}]{PhysRevLett.81.3108}%
  \BibitemOpen
  \bibfield  {author} {\bibinfo {author} {\bibfnamefont {D.}~\bibnamefont
  {Jaksch}}, \bibinfo {author} {\bibfnamefont {C.}~\bibnamefont {Bruder}},
  \bibinfo {author} {\bibfnamefont {J.~I.}\ \bibnamefont {Cirac}}, \bibinfo
  {author} {\bibfnamefont {C.~W.}\ \bibnamefont {Gardiner}}, \ and\ \bibinfo
  {author} {\bibfnamefont {P.}~\bibnamefont {Zoller}},\ }\href@noop {}
  {\bibfield  {journal} {\bibinfo  {journal} {Phys. Rev. Lett.}\ }\textbf
  {\bibinfo {volume} {81}},\ \bibinfo {pages} {3108} (\bibinfo {year}
  {1998})}\BibitemShut {NoStop}%
\bibitem [{\citenamefont {Pichler}\ \emph {et~al.}(2010)\citenamefont
  {Pichler}, \citenamefont {Daley},\ and\ \citenamefont
  {Zoller}}]{cold_dephasing}%
  \BibitemOpen
  \bibfield  {author} {\bibinfo {author} {\bibfnamefont {H.}~\bibnamefont
  {Pichler}}, \bibinfo {author} {\bibfnamefont {A.~J.}\ \bibnamefont {Daley}},
  \ and\ \bibinfo {author} {\bibfnamefont {P.}~\bibnamefont {Zoller}},\
  }\href@noop {} {\bibfield  {journal} {\bibinfo  {journal} {Phys. Rev. A}\
  }\textbf {\bibinfo {volume} {82}},\ \bibinfo {pages} {063605} (\bibinfo
  {year} {2010})}\BibitemShut {NoStop}%
\bibitem [{\citenamefont {Hartmann}(2016{\natexlab{b}})}]{Hartmann_2016}%
  \BibitemOpen
  \bibfield  {author} {\bibinfo {author} {\bibfnamefont {M.~J.}\ \bibnamefont
  {Hartmann}},\ }\href@noop {} {\bibfield  {journal} {\bibinfo  {journal}
  {Journal of Optics}\ }\textbf {\bibinfo {volume} {18}},\ \bibinfo {pages}
  {104005} (\bibinfo {year} {2016}{\natexlab{b}})}\BibitemShut {NoStop}%
\bibitem [{\citenamefont {Breuer}\ and\ \citenamefont
  {Petruccione}(2002)}]{BRE02}%
  \BibitemOpen
  \bibfield  {author} {\bibinfo {author} {\bibfnamefont {H.~P.}\ \bibnamefont
  {Breuer}}\ and\ \bibinfo {author} {\bibfnamefont {F.}~\bibnamefont
  {Petruccione}},\ }\href@noop {} {\emph {\bibinfo {title} {The theory of open
  quantum systems}}}\ (\bibinfo  {publisher} {Oxford University Press},\
  \bibinfo {address} {Great Clarendon Street},\ \bibinfo {year}
  {2002})\BibitemShut {NoStop}%
\bibitem [{\citenamefont {Zyczkowski}\ \emph {et~al.}(1998)\citenamefont
  {Zyczkowski}, \citenamefont {Horodecki}, \citenamefont {Sanpera},\ and\
  \citenamefont {Lewenstein}}]{negativity}%
  \BibitemOpen
  \bibfield  {author} {\bibinfo {author} {\bibfnamefont {K.}~\bibnamefont
  {Zyczkowski}}, \bibinfo {author} {\bibfnamefont {P.}~\bibnamefont
  {Horodecki}}, \bibinfo {author} {\bibfnamefont {A.}~\bibnamefont {Sanpera}},
  \ and\ \bibinfo {author} {\bibfnamefont {M.}~\bibnamefont {Lewenstein}},\
  }\href@noop {} {\bibfield  {journal} {\bibinfo  {journal} {Phys. Rev. A}\
  }\textbf {\bibinfo {volume} {58}},\ \bibinfo {pages} {883} (\bibinfo {year}
  {1998})}\BibitemShut {NoStop}%
\bibitem [{\citenamefont {Thew}\ \emph {et~al.}(2002)\citenamefont {Thew},
  \citenamefont {Nemoto}, \citenamefont {White},\ and\ \citenamefont
  {Munro}}]{tomography}%
  \BibitemOpen
  \bibfield  {author} {\bibinfo {author} {\bibfnamefont {R.~T.}\ \bibnamefont
  {Thew}}, \bibinfo {author} {\bibfnamefont {K.}~\bibnamefont {Nemoto}},
  \bibinfo {author} {\bibfnamefont {A.~G.}\ \bibnamefont {White}}, \ and\
  \bibinfo {author} {\bibfnamefont {W.~J.}\ \bibnamefont {Munro}},\ }\href@noop
  {} {\bibfield  {journal} {\bibinfo  {journal} {Phys. Rev. A}\ }\textbf
  {\bibinfo {volume} {66}},\ \bibinfo {pages} {012303} (\bibinfo {year}
  {2002})}\BibitemShut {NoStop}%
\bibitem [{Note1()}]{Note1}%
  \BibitemOpen
  \bibinfo {note} {The nonzero matrix elements of the Hermitian $\Lambda
  ^{(i)}$ matrices, such that $\Lambda ^{(i)}_{r,s} = (\Lambda
  ^{(i)}_{s,r})^{\star }$, are the following ones: $\Lambda ^{(0)}_{11} =
  \Lambda ^{(0)}_{22} =\Lambda ^{(0)}_{33} = 1$, $ \Lambda ^{(1)}_{12} = 1$,
  $\Lambda ^{(2)}_{12} = - i$, $\Lambda ^{(3)}_{11} = - \Lambda ^{(3)}_{22} =
  1$, $\Lambda ^{(4)}_{13} = 1$,$\Lambda ^{(5)}_{13} = -i$, $\Lambda
  ^{(6)}_{23} =1$, $\Lambda ^{(7)}_{23} = -i$, $\Lambda ^{(8)}_{22} =
  1/\protect \sqrt {3}$ and $\Lambda ^{(8)}_{33} = -2/\protect \sqrt
  {3}$.}\BibitemShut {Stop}%
\bibitem [{Note2()}]{Note2}%
  \BibitemOpen
  \bibinfo {note} {The eight generators $\Lambda ^{(i)}$ can be expressed in
  terms of the bosonic annihilation and creation operators. Namely: $\Lambda
  ^{(1)} =\protect \frac {1}{2}(b b^{\dagger 2}+b^{2} b^{\dagger })$, $\Lambda
  ^{(2)}=\protect \frac {i}{2}(b^2 b^{\dagger }-b b^{\dagger 2})$, $\Lambda
  ^{(3)}=\protect \frac {3}{4}b^{2} b^{\dagger 2}-\protect \frac {1}{2}b
  b^{\dagger }$, $\Lambda ^{(4)}=\protect \frac {1}{\protect \sqrt
  {2}}(b^{\dagger 2}+b^2)$, $\Lambda ^{(5)}=\protect \frac {i}{\protect \sqrt
  {2}}(b^2 - b^{\dagger 2})$, $\Lambda ^{(6)}=\protect \frac {1}{\protect \sqrt
  {2}}(b^{\dagger 2} b+b^{\dagger } b^2)$, $\Lambda ^{(7)}=\protect \frac
  {i}{\protect \sqrt {2}}(b^{\dagger } b^{2} -b^{\dagger 2} b)$, and $\Lambda
  ^{(8)} =\protect \frac {1}{\protect \sqrt {3}}(bb^{\dagger }-b^{\dagger }
  b)$.}\BibitemShut {Stop}%
\bibitem [{\citenamefont {Besse}\ \emph {et~al.}(2020)\citenamefont {Besse},
  \citenamefont {Gasparinetti}, \citenamefont {Collodo}, \citenamefont
  {Walter}, \citenamefont {Remm}, \citenamefont {Krause}, \citenamefont
  {Eichler},\ and\ \citenamefont {Wallraff}}]{walraff}%
  \BibitemOpen
  \bibfield  {author} {\bibinfo {author} {\bibfnamefont {J.-C.}\ \bibnamefont
  {Besse}}, \bibinfo {author} {\bibfnamefont {S.}~\bibnamefont {Gasparinetti}},
  \bibinfo {author} {\bibfnamefont {M.~C.}\ \bibnamefont {Collodo}}, \bibinfo
  {author} {\bibfnamefont {T.}~\bibnamefont {Walter}}, \bibinfo {author}
  {\bibfnamefont {A.}~\bibnamefont {Remm}}, \bibinfo {author} {\bibfnamefont
  {J.}~\bibnamefont {Krause}}, \bibinfo {author} {\bibfnamefont
  {C.}~\bibnamefont {Eichler}}, \ and\ \bibinfo {author} {\bibfnamefont
  {A.}~\bibnamefont {Wallraff}},\ }\href@noop {} {\bibfield  {journal}
  {\bibinfo  {journal} {Phys. Rev. X}\ }\textbf {\bibinfo {volume} {10}},\
  \bibinfo {pages} {011046} (\bibinfo {year} {2020})}\BibitemShut {NoStop}%
\bibitem [{\citenamefont {Zwolak}\ and\ \citenamefont {Vidal}(2004)}]{mpo2}%
  \BibitemOpen
  \bibfield  {author} {\bibinfo {author} {\bibfnamefont {M.}~\bibnamefont
  {Zwolak}}\ and\ \bibinfo {author} {\bibfnamefont {G.}~\bibnamefont {Vidal}},\
  }\href@noop {} {\bibfield  {journal} {\bibinfo  {journal} {Phys. Rev. Lett.}\
  }\textbf {\bibinfo {volume} {93}},\ \bibinfo {pages} {207205} (\bibinfo
  {year} {2004})}\BibitemShut {NoStop}%
\bibitem [{\citenamefont {Verstraete}\ \emph {et~al.}(2004)\citenamefont
  {Verstraete}, \citenamefont {Garc\'{\i}a-Ripoll},\ and\ \citenamefont
  {Cirac}}]{mpo3}%
  \BibitemOpen
  \bibfield  {author} {\bibinfo {author} {\bibfnamefont {F.}~\bibnamefont
  {Verstraete}}, \bibinfo {author} {\bibfnamefont {J.~J.}\ \bibnamefont
  {Garc\'{\i}a-Ripoll}}, \ and\ \bibinfo {author} {\bibfnamefont {J.~I.}\
  \bibnamefont {Cirac}},\ }\href@noop {} {\bibfield  {journal} {\bibinfo
  {journal} {Phys. Rev. Lett.}\ }\textbf {\bibinfo {volume} {93}},\ \bibinfo
  {pages} {207204} (\bibinfo {year} {2004})}\BibitemShut {NoStop}%
\bibitem [{Note3()}]{Note3}%
  \BibitemOpen
  \bibinfo {note} {Previous studies of two-qubit systems with non-Markovian
  environments have revealed entanglement revival effects \cite {revival,
  revival2}. In our system, the two-site dynamics is non-unitary and
  non-Markovian even for $\gamma = 0$ since the other sites of the chain have
  been traced out for the calculation of the negativity.}\BibitemShut {Stop}%
\bibitem [{\citenamefont {Ma}\ \emph {et~al.}(2017)\citenamefont {Ma},
  \citenamefont {Owens}, \citenamefont {Houck}, \citenamefont {Schuster},\ and\
  \citenamefont {Simon}}]{stabilizer}%
  \BibitemOpen
  \bibfield  {author} {\bibinfo {author} {\bibfnamefont {R.}~\bibnamefont
  {Ma}}, \bibinfo {author} {\bibfnamefont {C.}~\bibnamefont {Owens}}, \bibinfo
  {author} {\bibfnamefont {A.}~\bibnamefont {Houck}}, \bibinfo {author}
  {\bibfnamefont {D.~I.}\ \bibnamefont {Schuster}}, \ and\ \bibinfo {author}
  {\bibfnamefont {J.}~\bibnamefont {Simon}},\ }\href@noop {} {\bibfield
  {journal} {\bibinfo  {journal} {Phys. Rev. A}\ }\textbf {\bibinfo {volume}
  {95}},\ \bibinfo {pages} {043811} (\bibinfo {year} {2017})}\BibitemShut
  {NoStop}%
\bibitem [{\citenamefont {Bellomo}\ \emph {et~al.}(2007)\citenamefont
  {Bellomo}, \citenamefont {Lo~Franco},\ and\ \citenamefont
  {Compagno}}]{revival}%
  \BibitemOpen
  \bibfield  {author} {\bibinfo {author} {\bibfnamefont {B.}~\bibnamefont
  {Bellomo}}, \bibinfo {author} {\bibfnamefont {R.}~\bibnamefont {Lo~Franco}},
  \ and\ \bibinfo {author} {\bibfnamefont {G.}~\bibnamefont {Compagno}},\
  }\href@noop {} {\bibfield  {journal} {\bibinfo  {journal} {Phys. Rev. Lett.}\
  }\textbf {\bibinfo {volume} {99}},\ \bibinfo {pages} {160502} (\bibinfo
  {year} {2007})}\BibitemShut {NoStop}%
\bibitem [{\citenamefont {Ficek}\ and\ \citenamefont
  {Tana\ifmmode~\acute{s}\else \'{s}\fi{}}(2006)}]{revival2}%
  \BibitemOpen
  \bibfield  {author} {\bibinfo {author} {\bibfnamefont {Z.}~\bibnamefont
  {Ficek}}\ and\ \bibinfo {author} {\bibfnamefont {R.}~\bibnamefont
  {Tana\ifmmode~\acute{s}\else \'{s}\fi{}}},\ }\href@noop {} {\bibfield
  {journal} {\bibinfo  {journal} {Phys. Rev. A}\ }\textbf {\bibinfo {volume}
  {74}},\ \bibinfo {pages} {024304} (\bibinfo {year} {2006})}\BibitemShut
  {NoStop}%
\end{thebibliography}%
\bibliographystyle{apsrev4-1}
\clearpage

\end{document}